\documentclass[reprint,amsmath,amssymb,aps,pra,superscriptaddress]{revtex4-1}
\usepackage[utf8]{inputenc}
\usepackage[T1]{fontenc}
\usepackage{amsmath}
\usepackage{amsthm}
\usepackage{amssymb}
\usepackage{braket}
\usepackage{graphicx}
\usepackage{bm}
\usepackage{url}
\usepackage{hyperref}
\usepackage{dcolumn}
\usepackage{complexity}
\usepackage{lmodern}
\usepackage{natbib}
\usepackage{color}
\usepackage{booktabs}
\usepackage{epstopdf}
\usepackage{siunitx}
\usepackage{subfigure}
\usepackage{comment}
\usepackage{cancel}
\usepackage{resizegather}

\begin{document}
\title{Controlling the Multiport Nature of Bragg Diffraction in Atom Interferometry}
\author{Richard H Parker}
\author{Chenghui Yu}
\author{Brian Estey}
\author{Weicheng Zhong}
\affiliation{University of California, Berkeley, CA 94720, United States}
\author{Eric Huang}
\affiliation{The University of Sydney, NSW 2006, Australia}
\author{Holger Müller}
\affiliation{University of California, Berkeley, CA 94720, United States}

\date{\today}

\begin{abstract}
Bragg diffraction has been used in atom interferometers because it allows signal enhancement through multiphoton momentum transfer and suppression of systematics by not changing the internal state of atoms. Its multi-port nature, however, can lead to parasitic interferometers, allows for intensity-dependent phase shifts in the primary interferometers, and distorts the ellipses used for phase extraction. We study and suppress these unwanted effects. Specifically, phase extraction by ellipse fitting and the resulting systematic phase shifts are calculated by Monte Carlo simulations. Phase shifts arising from the thermal motion of the atoms are controlled by spatial selection of atoms and an appropriate choice of Bragg intensity. In these simulations, we found that Gaussian Bragg pulse shapes yield the smallest systematic shifts. Parasitic interferometers are suppressed by a ``magic'' Bragg pulse duration. The sensitivity of the apparatus was improved by the addition of AC Stark shift compensation, which permits direct experimental study of sub-part-per-billion (ppb) systematics. This upgrade allows for a 310\,$\hbar k$ momentum transfer, giving an unprecedented 6.6\,Mrad measured in a Ramsey-Bordé interferometer. 
\end{abstract}
\maketitle

Atom interferometers have been used for tests of fundamental physics such as the isotropy of gravity \cite{PhysRevD.80.016002}, the equivalence principle \cite{PhysRevLett.113.023005, PhysRevLett.115.013004, PhysRevLett.117.023001,2015arXiv150705820S}, the search for dark-sector particles \cite{Hamilton849, 1603.06587}, and measurements of the fine structure constant $\alpha$ \cite{PhysRevLett.106.080801, PhysRevA.90.063606}, which characterizes the strength of the electromagnetic interaction. This constant can be obtained from the electron's gyromagnetic anomaly $g_{e}-2$. At the current accuracy, this involves > 10,000 Feynman diagrams, as well as muonic and hadronic physics \cite{PhysRevLett.109.111807}. At increased accuracy, the tauon and the weak interaction will also be included. Since this path leads to 0.24\,ppb accuracy \cite{PhysRevLett.100.120801}, an independent measurement of $\alpha$ would create a unique test for the standard model. The best such measurements of $\alpha$ are currently based on the recoil energy $\hbar^{2}k^{2}/2m_{\mathrm{At}}$ of an atom of mass $m_{\mathrm{At}}$ that has scattered a photon of momentum $\hbar k$ \cite{0605125v1, 1402-4896-2002-T102-014}. This measurement yields $\hbar / m_{\mathrm{At}}$, and yields $\alpha$ to 0.66\,ppb \cite{PhysRevLett.106.080801} via the relation
\begin{align}
    \alpha^{2} = \frac{2 R_{\infty}}{c} \frac{m_{At}}{m_{e}} \frac{\hbar}{m_{At}}. \label{eqn:alpha}
\end{align}
The Rydberg constant $R_{\infty}$ is known to 0.005\,ppb accuracy, and the atom-to-electron mass ratio is known to better than 0.1\,ppb for many species \cite{Sturm2014}. 

In this paper, we improve the accuracy of a measurement of the fine structure constant using Bragg diffraction, by both increasing the sensitivity of the experiment and a thorough theoretical analysis of important systematic effects. In Section I, we present an enhancement in the sensitivity of an atom interferometer (AI) by AC Stark compensation, which allows faster integration. In Section II, we investigate aberrations to the elliptical shape used for phase extraction which arise from the diffraction phase. Section III shows how this leads to phase shifts due to thermal motion, and Section IV describes how spatial filtering can be used to suppress those shifts. In Section V, we consider the influence of the Bragg pulse shape, and in Section VI we examine the influence of parasitic AIs (as well as find a ``magic'' duration which can be used to suppress them). 

\section{Improved Sensitivity through stark compensation}

\begin{figure}[h]
	\centering
	\includegraphics[width=\columnwidth]{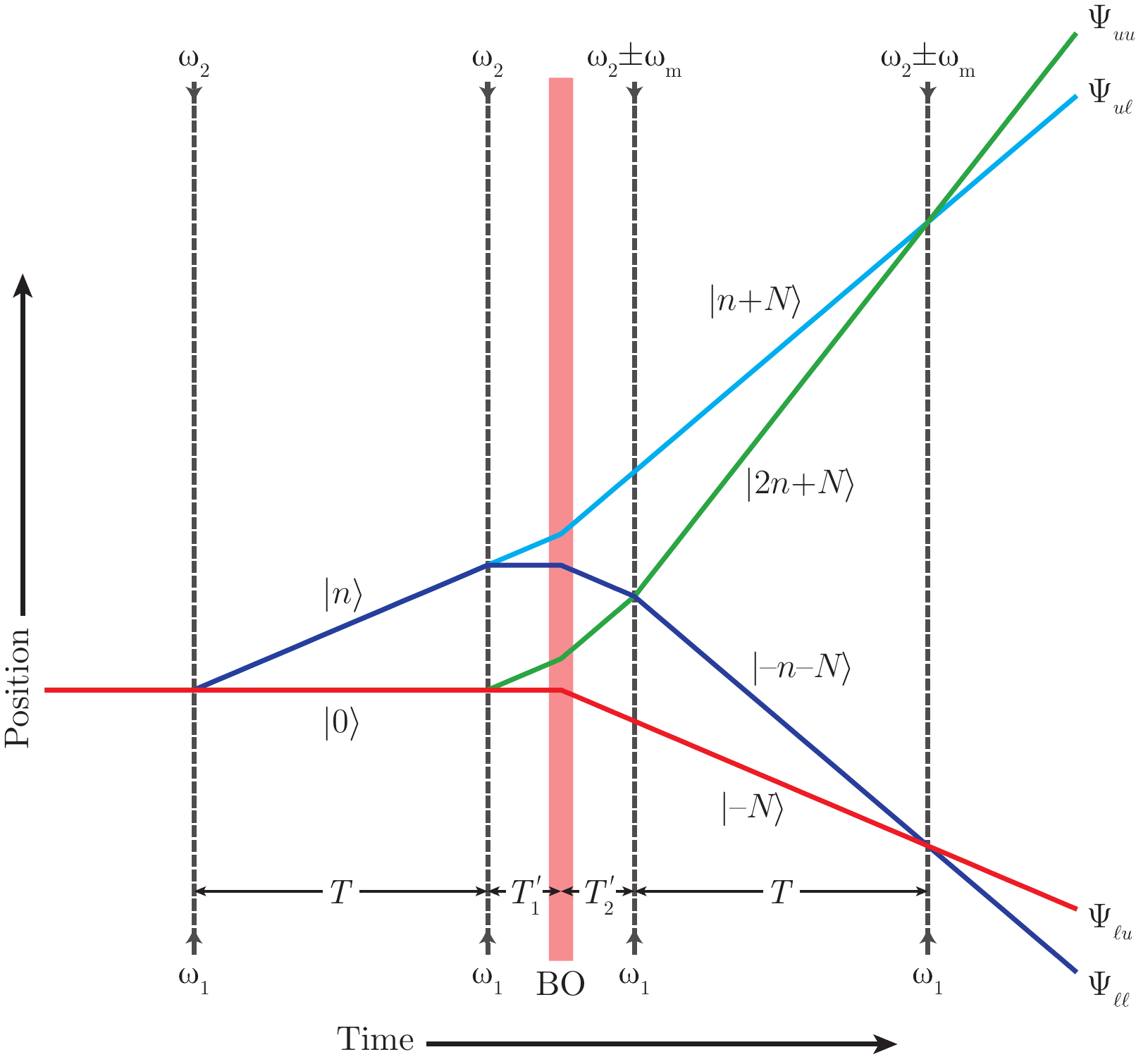}
	\caption{Geometry of the two simultaneous conjugate Ramsey-Bordé interferometers. The four Bragg pulses that form the interferometers are shown in dashed lines; Bloch Oscillations (BO) between the 2nd and 3rd Bragg pulses provide additional momentum splitting.}
	\label{fig:rbmomentum}
\end{figure}

The atom interferometer discussed in this paper has been described in detail in Ref \cite{PhysRevLett.115.083002}; the geometry is shown in Figure \ref{fig:rbmomentum}. Two cesium Ramsey-Bordé interferometers are operated in a simultaneous-conjugate configuration \cite{PhysRevLett.103.050402}, with each $2n$-photon beamsplitter formed by a Bragg pulse that splits the atoms by a total of 2$n\hbar k$, where $\hbar k$ is the photon momentum, without changing the internal state of the atoms. A frequency ramp that accelerates the lattice, driving Bloch oscillations, is applied in the middle of the sequence, to provide additional momentum splitting by $2N\hbar k$. Up-going and down-going ramps are applied simultaneously. The total phase difference between the two interferometers is
\begin{align}
    \Delta\Phi = 16n(n+N)\omega_r T - 2n\omega_m T.\label{eqn:01}
\end{align}
where $T$ is the separation time between the first and second laser pulses (also equal to that between the third and fourth), 2$\omega_{m}$ is the detuning between the frequencies used for the final two pulses, and $\omega_r=\hbar k^2/2m$ is the recoil frequency we seek to measure. Increasing the Bragg order, Bloch order, or pulse separation time $T$ will increase the total accumulated phase, and therefore the experiment's sensitivity. 

As the pulse separation time is increased, distortions in the wavefronts of the Bragg and Bloch beams (which originate from the same fiber port) will result in spatially-varying AC Stark shifts that lead to decoherence \cite{Kovachy2015}. The distortions can be caused by diffraction from obstructions such as dust on optics/viewports, the circular aperture of the fiber port, and the inner wall of the vacuum system. 

\begin{figure}[h]
	\centering
	\includegraphics[width=\columnwidth]{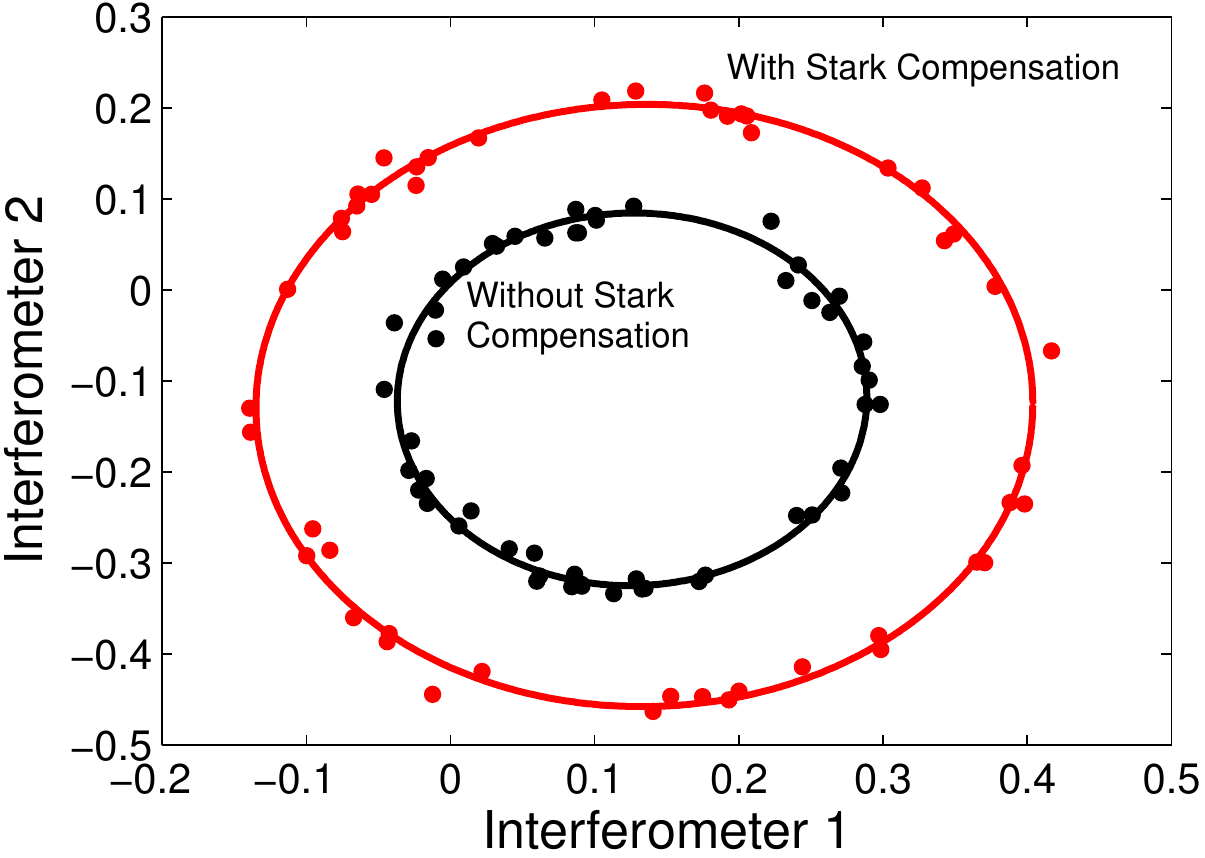}
	\caption{Comparison of ellipses with and without Stark compensation. Data taken with $n$=5, $N$=25. The solid lines are ellipse fits.}
	\label{fig:comp}
\end{figure}

To suppress this effect, we apply a beam from the same fiber port as the Bragg and Bloch beams, with the same intensity but the opposite single-photon detuning, as suggested in  \cite{Kovachy2015}. This beam is single-frequency, does not satisfy Bragg resonance, and does not drive Bragg transitions. This beam compensates for the variable AC Stark shift and enhances the Bloch order and pulse separation time at which we have acceptable contrast (see Figure \ref{fig:comp}). 

\begin{figure}[h]
	\centering
	\includegraphics[width=\columnwidth]{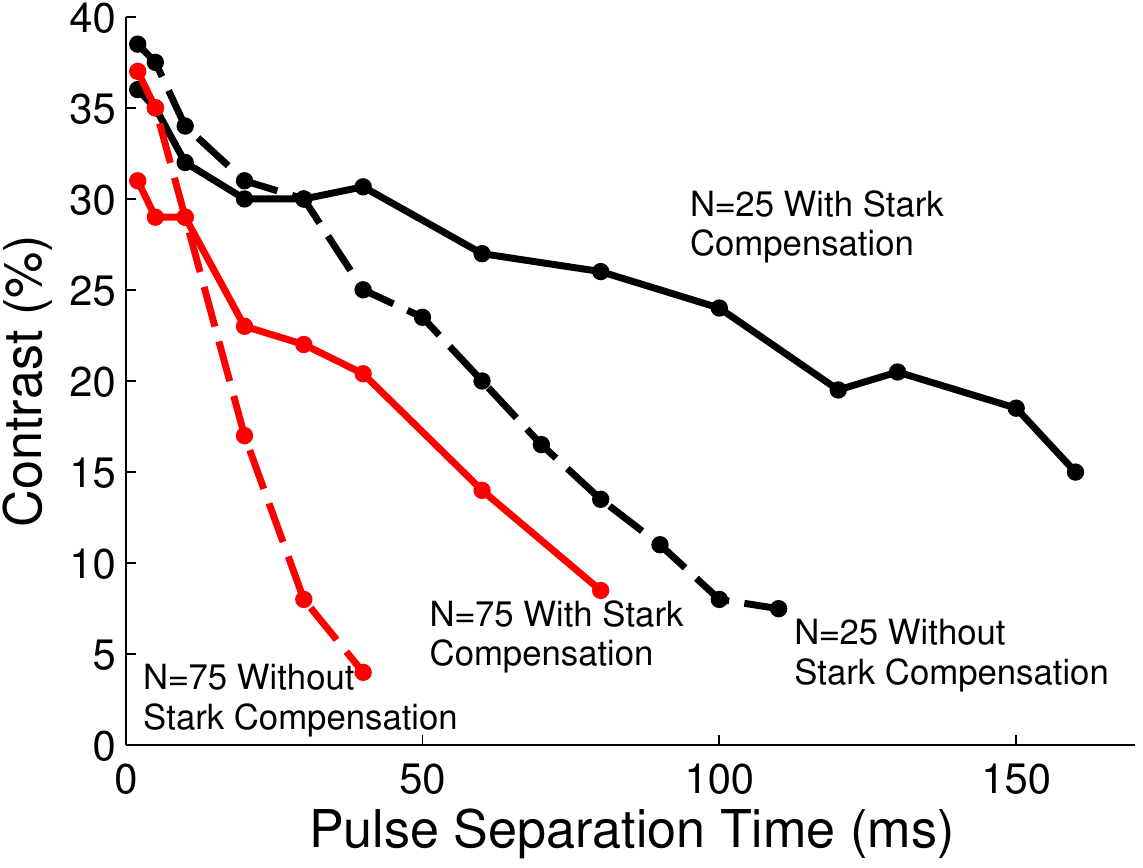}
	\caption{Data taken with $N$=25 and $N$=75 ($n$=5) showing the enhancement of contrast resulting from the addition of the Stark compensation beam.}
	\label{fig:acssc}
\end{figure}

Without this upgrade, $N$=25 was the largest Bloch order at which usable ellipses could be observed, with a pulse separation time of $T$=80\,ms. Stark compensation allows coherence to be observed with $N$=75 up to a maximum pulse separation time of $T$=80\,ms. Beyond this point, other decoherence mechanisms (including thermal expansion of the atom cloud and single-photon scattering) dominate, as shown in Figure \ref{fig:acssc}. The momentum splitting between the ``fastest'' and ``slowest'' arms of the interferometer is $2(n+2N)=310\hbar k$ (1.1\,m/s relative motion), giving an unprecedented 6.6\,Mrad measured in a Ramsey-Bordé interferometer. This represents the largest measured phase of any Ramsey-Bordé interferometer. Not only does this upgrade allow a measurement of $\alpha$ with a higher integration rate, it also permits the study of sub-ppb systematic effects.

\section{Diffraction phase extraction via ellipse fitting} \label{theory}

Extracting the differential and common-mode phases from a pair of conjugate interferometers by ellipse fitting is well-established \cite{Foster:02}. However, the use of Bragg (i.e. multi-port) beamsplitters will distort the ellipses, and could potentially produce a systematic error in the ellipse fit. Here we consider the validity of ellipse fitting in the case of multi-port beamsplitters, by extending the analysis in Ref \cite{PhysRevLett.115.083002}. We begin with the optical Bloch equations in the cases of a single-frequency and a multi-frequency Bragg pulse \cite{PhysRevLett.115.083002}:

\begin{gather}
\label{eq:braggsystem}
i \dot{g}_n = 2\Omega \left[ g_{n+1} e^{i2 \delta t} e^{-i4 \left( 2n+1 \right) \omega_r t} + g_{n-1} e^{-i2 \delta t} e^{i4 \left( 2n-1 \right) \omega_r t} \right], \\
\label{eq:multisystem1}
i \dot{g_n} = 4 \cos (\omega_m t) \Omega \left[ g_{n+1} e^{i2 \delta t} e^{-i4 \left( 2n+1 \right) \omega_r t} + g_{n-1} e^{-i2 \delta t} e^{i4 \left( 2n-1 \right) \omega_r t} \right].
\end{gather}

Each is an infinite set of differential equations, where $g_{n}(z,t)$ are plane-wave momentum states of the atomic ground state, $\Omega$ is the two-photon Rabi frequency, and 2$\delta = \omega_{1} - \omega_{2}$ is the frequency difference between the counter-propagating field and the average of the co-propagating fields. 

Combining \eqref{eq:braggsystem} and \eqref{eq:multisystem1}, we can calculate the total phase shift in the interferometer due to the beam splitter diffraction phase. At each beam splitter along a particular path, the atom acquires a diffraction phase \cite{Lan554,PhysRevA.88.053608}.  There are eight relevant paths. For each of the two interferometers (upper and lower), there are two possible trajectories (upper and lower), and two output ports for each of those trajectories (upper and lower) for a total of $2^3$ paths (see Figure \ref{fig:rbmomentum}). 

We evaluate $\left< b \right| \hat{H}_n \left| a \right>$, the amplitude for the Bragg pulse to transfer an atom from momentum state $a$ moving at $2 a \hbar k$ into a momentum state $b$ when driven with laser frequencies at $\delta=4n\omega_r$, by numerically solving Eqn. \ref{eq:braggsystem}. Similarly, the matrix elements $\left< b \right| \hat{H}_{n,N} \left| a \right>$ can be found by integrating the multi-frequency optical Bloch equation for $\delta=4n\omega_r$ and $\omega_m=8(n+N)\omega_r$.

The matrix elements for $\hat{H}_n$ and $\hat{H}_{n,N}$ are symmetric, as well as invariant under a momentum transformation \cite{PhysRevLett.115.083002}:

\begin{align}
\label{eq:diffractionsym}
	\left< b \right| \hat{H}_{n(,N)} \left| a \right> &= \left< a \right| \hat{H}_{n(,N)} \left| b \right> \\
	\nonumber
	&=\left< b+c \right| \hat{H}_{n+c(,N)} \left| a+c \right>.
\end{align}

\noindent
For the two-frequency Hamilton $H_n$, there is also symmetry about the Bragg resonance condition. For any value  $\{a,b,c\}\in\mathbb{R}$, then

\[
	\left< n + b \right| \hat{H}_{2n+c} \left| n + a \right> = \left< n - b \right| \hat{H}_{2n-c} \left| n - a \right>.
\]

\noindent
Additionally, for large modulation frequency $\omega_m$, the multi-frequency Hamiltonian $\hat{H}_{n,N}$ looks like a momentum shifted two-frequency Hamiltonian $\hat{H}_n$:
\begin{equation}
\begin{aligned}
\label{eq:undiffractedsym}
	\left< b \right| \hat{H}_{n} \left| a \right> &= \lim_{N\rightarrow\infty} \left< b + n+N \right| \hat{H}_{n,N} \left| a+n+N \right> \\
	&= \lim_{N\rightarrow\infty} \left< b-n-N \right| \hat{H}_{n,N} \left| a-n-N \right>.
\end{aligned}
\end{equation}

We write the complex amplitude along an interferometer path as $c_{ijk}$, where $i$ determines the interferometer, $j$ determines the path within that interferometer, and $k$ determines the output port. Each index can take two values, $u$ or $\ell$, signifying ``upper'' and ``lower'' respectively.  

\begin{align*}
c_{\ell u \ell} &= \left< -n-N \right| \hat{H}_{n,N} \left| -n-N \right> \left< -N \right| \hat{H}_{n,N} \left| -n-N \right> \\ & \phantom{= \left< -n-N \right| \hat{H}_{n,N} \left| -n-N \right>} \times \left< n \right| \hat{H}_n \left| 0 \right>^2, \\
c_{\ell \ell \ell} &= \left< -n-N \right| \hat{H}_{n,N} \left| -N \right> \left< -N \right| \hat{H}_{n,N} \left| -N \right>\left< 0 \right| \hat{H}_n \left| 0 \right>^2 , \\
c_{\ell u u} &= \left< -N \right| \hat{H}_{n,N} \left| -n-N \right>^2\left< n \right| \hat{H}_n \left| 0 \right>^2, \\
c_{\ell \ell u} &= \left< -N \right| \hat{H}_{n,N} \left| -N \right>^2\left< 0 \right| \hat{H}_n \left| 0 \right>^2, \\
c_{uu\ell} &= \left< n+N \right| \hat{H}_{n,N} \left| n+N \right>^2 \left< n \right| \hat{H}_n \left| n \right>\left< n \right| \hat{H}_n \left| 0 \right>, \\
c_{u\ell \ell} &= \left< n+N \right| \hat{H}_{n,N} \left| 2n+N \right>^2 \left< n \right| \hat{H}_n \left| 0 \right>\left< 0 \right| \hat{H}_n \left| 0 \right>, \\
c_{uuu} &= \left< 2n+N \right| \hat{H}_{n,N} \left| n+N \right> \left< n+N \right| \hat{H}_{n,N} \left| n+N \right>  \\ & \phantom{=\left< 2n+N \right| \hat{H}_{n,N} \left| n+N \right>} \times \left< n \right| \hat{H}_n \left| n \right>\left< n \right| \hat{H}_n \left| 0 \right>, \\
c_{u\ell u} &=  \left< 2n+N \right| \hat{H}_{n,N} \left| 2n+N \right> \left< n+N \right| \hat{H}_{n,N} \left| 2n+N \right> \\ & \phantom{=\left< 2n+N \right| \hat{H}_{n,N} \left| 2n+N \right>} \times \left< n \right| \hat{H}_n \left| 0 \right>\left< 0 \right| \hat{H}_n \left| 0 \right>.
\end{align*}

We define the phases along the lower and upper trajectories to be $\phi_\ell=\phi_c+\phi_d$ and $\phi_u=\phi_c-\phi_d$, where $\phi_d$ is the differential phase to be measured and $\phi_c$ is a common mode term which fluctuates between zero and $2\pi$ due to vibrations. The interferometer outputs are then

\begin{align*}
\Psi_{\ell \ell} &= \left|-n-N\right> \left[ c_{\ell u \ell}+e^{i\phi_\ell}c_{\ell \ell \ell} \right], \\
\Psi_{\ell u} &= \left|-N\right> \left[ c_{\ell u u}+e^{i\phi_\ell}c_{\ell \ell u} \right], \\
\Psi_{u\ell} &= \left|n+N\right> \left[ c_{uu\ell}+e^{i\phi_u}c_{u\ell \ell} \right], \\
\Psi_{uu} &= \left|2n+N\right> \left[ c_{uuu}+e^{i\phi_u}c_{u\ell u} \right],
\end{align*}

\noindent
where the individual beam-splitter phases $\phi_{q}$  ($\phi_{\ell\ell u},\phi_{\ell u \ell},\dots$) are given by $c_{q}=\left|c_{q}\right|e^{i\phi_{q}}$. Assuming $\left|c_{i}\right| = \left|c_{j}\right|$, the measured populations of the lower interferometer output ports are 

\begin{align*}
	\left|\Psi_{\ell \ell}\right|^2 &=\cos^2\left(\frac{1}{2}\left(\phi_\ell + \phi_{\ell \ell \ell}-\phi_{\ell u \ell}\right)\right), \\
	\left|\Psi_{\ell u}\right|^2 &=\sin^2\left(\frac{1}{2}\left(\phi_\ell + \phi_{\ell \ell u}-\phi_{\ell u u}-\pi\right)\right),
\end{align*}

\noindent
where we have inserted a $\pi$ phase shift by hand in the last equation to produce the usual $\sin^2$ interference result. By applying the substitution $\Phi_\ell=\phi_\ell + \phi_{\ell \ell \ell} - \phi_{\ell u \ell}$, the amplitudes of the lower output port simplify to

\begin{align*}
	\left|\Psi_{\ell u}\right|^2 &=\sin^2\left(\frac{1}{2}\left(\Phi_\ell + (\phi_{\ell \ell u}-\phi_{\ell u u}) - (\phi_{\ell \ell \ell}-\phi_{\ell u \ell}) -\pi\right)\right)\\
	&= \sin^2\left(\frac{1}{2}\left(\Phi_\ell + \Delta \phi_\ell \right) \right),
\end{align*}

\noindent
where we have introduced $\Delta \phi_\ell = (\phi_{\ell \ell u}-\phi_{\ell u u}) - (\phi_{\ell \ell \ell}-\phi_{\ell u \ell}) -\pi$ as the difference in diffraction phase between the two output ports of the lower interferometer. Since the beam-splitter phases also encode the usual $\pi/2$ phase shifts intrinsic to beam-splitters, the aforementioned $\pi$ phase shift is included to make $\Delta \phi_\ell$ small.

Plotting $\left| \Psi_{\ell \ell} \right|^2$ vs. $\left| \Psi_{u \ell} \right|^2$ while the common-mode phase fluctuates creates an ellipse, whose shape allows the determination of the differential phase $\phi_d$. Although it is possible to estimate the phase difference between the upper and lower interferometer by fitting the ellipse $\left| \Psi_{\ell \ell} \right|^2$ vs. $\left| \Psi_{u \ell} \right|^2$, this method is sensitive to fluctuation in the absolute number of atoms. A more robust technique is to use the normalized difference between the interferometer outputs $(\left| \Psi_{\ell \ell} \right|^2 - \left| \Psi_{\ell u} \right|^2)/(\left| \Psi_{\ell \ell} \right|^2+\left| \Psi_{\ell u} \right|^2)$ and  $(\left| \Psi_{u \ell} \right|^2 - \left| \Psi_{u u} \right|^2)/(\left| \Psi_{u \ell} \right|^2+\left| \Psi_{u u} \right|^2)$. For an ideal interferometer with output populations $\left| \Psi_{\ell} \right|^2 = A\cos(\phi/2)^2$ and $\left| \Psi_{u} \right|^2 = A \sin(\phi/2)^2$, the normalized difference reduces to $(\left| \Psi_{\ell} \right|^2 - \left| \Psi_{u} \right|^2)/(\left| \Psi_{\ell} \right|^2+\left| \Psi_{u} \right|^2)=\cos(\phi)$, independent of the signal amplitude $A$.

\begin{figure}[h]
	\centering
	\includegraphics[width=\columnwidth]{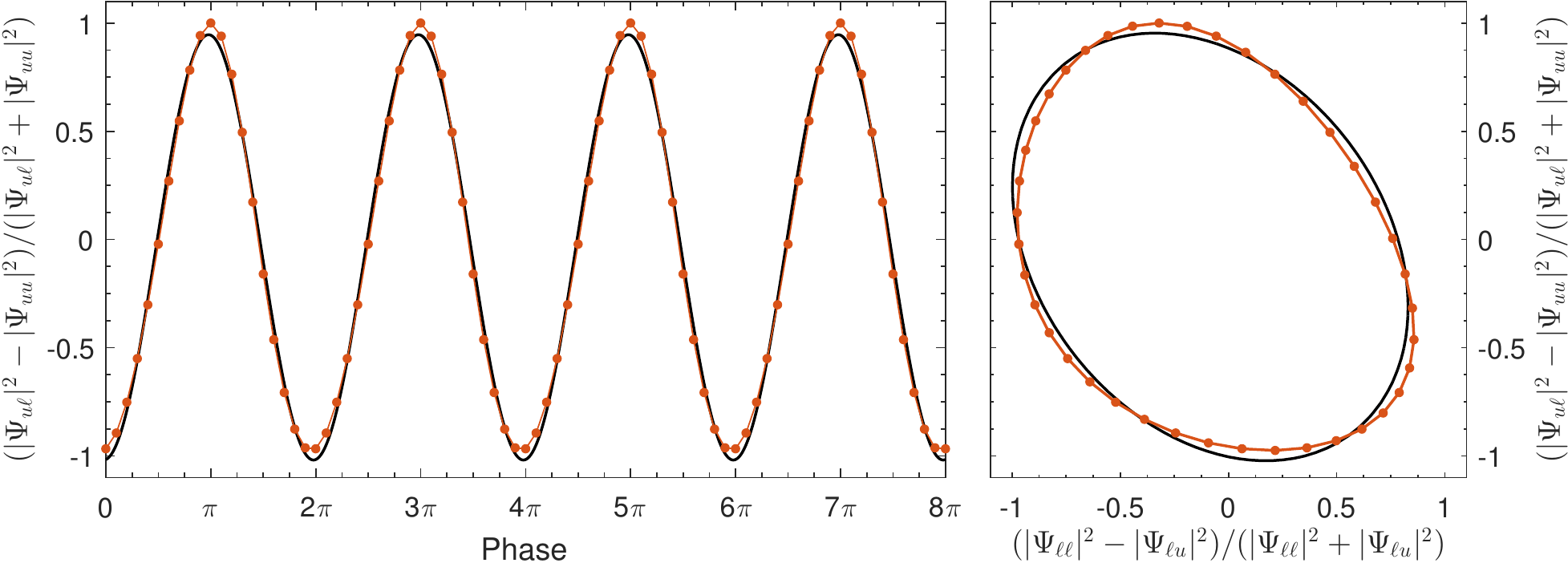}
	\caption{Non-ideal ellipse for parameters $n=5$ $N=0$ $\Omega_0= 7.81 \omega_r$, and a detuning from Bragg resonance $\delta v=0.3 v_r$ (where $v_r$ is the recoil velocity). The red curve includes the three terms in the RHS of Eq. \eqref{eq:ellipseaprox}; the black curve includes the first two.}
	\label{fig:nonellipse}
\end{figure}

However, in the case of Bragg diffraction, the sum of the interfering populations is not unity, $\left| \Psi_{\ell \ell} \right|^2 + \left| \Psi_{\ell u} \right|^2\neq1$, due to beam splitter losses. The effect of this can be seen in Fig. \ref{fig:nonellipse} where the parameters are chosen to be far from the ideal case of an interferometer formed by two-port beam-splitters; in this case the normalized outputs of the two conjugate interferometers don't form an ellipse.

\noindent
If the difference between the diffraction phase of the two output ports $\Delta \phi_{\ell}$ is small, then we can approximate the normalized population as
\begin{equation}
\begin{gathered}
\frac{\left| \Psi_{\ell \ell}\right|^2 - \left| \Psi_{\ell u}\right|^2}{\left| \Psi_{\ell \ell}\right|^2 + \left| \Psi_{\ell u}\right|^2} \approx \cos\left(\Phi_\ell\right) - \frac{1}{2} \Delta\phi_\ell \sin\left(\Phi_\ell \right) \\ + \frac{1}{4} \Delta \phi_\ell  \sin\left(2\Phi_\ell\right) + \mathcal{O}(\Delta\phi_{\ell}^2).
\end{gathered}\label{eq:ellipseaprox}
\end{equation}

\noindent
Recall that the phase $\phi_\ell$ (and therefore $\Phi_\ell$) contains a common mode term $\phi_{c}$ which fluctuates between zero and $2\pi$ due to vibrations. Interpreting $\phi_c$ as a parametric variable, we can further approximate \eqref{eq:ellipseaprox} by ignoring the second perturbative term that has twice the frequency in $\Phi_\ell$ (and therefore in $\phi_c$) as it will tend to average out when fitting \eqref{eq:ellipseaprox} to the $x/y$ component of an ellipse with the functional form $A\cos(\phi_c+\phi)+B$. Utilizing a Taylor-series approximation, the normalized population for the lower interferometer can be simplified to

\begin{align*}
x= \frac{\left| \Psi_{\ell \ell}\right|^2 - \left| \Psi_{\ell u}\right|^2}{\left| \Psi_{\ell \ell}\right|^2 + \left| \Psi_{\ell u}\right|^2} &\approx \cos\left(\Phi_\ell\right) - \frac{1}{2} \Delta\phi_\ell \sin\left(\Phi_\ell \right) \\
&\approx \cos\left(\Phi_\ell + \frac{\Delta\phi_\ell}{2}\right),
\end{align*}

\noindent
and similarly for the upper interferometer,

\begin{align*}
	y= \frac{\left| \Psi_{u \ell}\right|^2 - \left| \Psi_{u u}\right|^2}{\left| \Psi_{u \ell}\right|^2 + \left| \Psi_{u u}\right|^2} \approx \cos\left(\Phi_u + \frac{\Delta \phi_u}{2}\right),
\end{align*}

\noindent
where $\Phi_u=\phi_u + \phi_{u \ell \ell} - \phi_{u u \ell}$ and $\Delta \phi_u = (\phi_{udu} -\phi_{uuu}) -(\phi_{udd} -\phi_{uud})-\pi$. The extracted differential phase after ellipse fitting $(x,y)$ is then

\begin{align}
	\nonumber
	\Delta\varphi &= (\Phi_\ell+\frac{\Delta\phi_\ell}{2}) - (\Phi_u+\frac{\Delta\phi_u}{2}) \\
	\label{eq:diffractionpairs}
	&= 2\phi_d + \frac{1}{2}\left[(\phi_{\ell\ell\ell}-\phi_{u \ell \ell})-(\phi_{\ell u\ell}-\phi_{uu\ell})\right] \\ &\phantom{{}= {2\phi_d}}  + \nonumber \frac{1}{2}\left[(\phi_{\ell\ell u}-\phi_{u \ell u})-(\phi_{\ell u u}-\phi_{uuu})\right] \\ \nonumber
	&= 2\phi_d + \frac{1}{2}\left[(\phi_{\ell\ell\ell}-\phi_{\ell u\ell})-(\phi_{u \ell\ell}-\phi_{uu\ell})\right] \\ &\phantom{{}= {2\phi_d}}+ \nonumber\frac{1}{2}\left[(\phi_{\ell\ell u}-\phi_{\ell u u})-(\phi_{u \ell u}-\phi_{uuu})\right] \\ \nonumber
	\nonumber
	&= 2\phi_d+\phi_0
\end{align}

\noindent
where $\phi_0$ is the measured diffraction phase systematic due to beam-splitter losses. Therefore to leading order the differential diffraction phases for different output ports do not produce a systematic effect in a measurement of the fine structure constant. 

We can also take advantage of certain cancellations to determine which pulses contribute most to the measured diffraction phase. Expanding each diffraction phase pair in \eqref{eq:diffractionpairs} in terms of the Hamiltonian matrix elements,

\begin{widetext}
\begin{align}
	\phi_{\ell\ell\ell}-\phi_{u\ell\ell}&= \arg\left(\frac{\left< -n-N \right| \hat{H}_{n,N} \left| -N \right> \left< -N \right| \hat{H}_{n,N} \left| -N \right>\left< 0 \right| \hat{H}_n \left| 0 \right>^2}{\left< n+N \right| \hat{H}_{n,N} \left| 2n+N \right>^2 \left< n \right| \hat{H}_n \left| 0 \right>\left< 0 \right| \hat{H}_n \left| 0 \right>}\right) \label{eq9} \\ 
	\phi_{\ell u\ell}-\phi_{u u\ell}&= \arg\left(\frac{\left< -n-N \right| \hat{H}_{n,N} \left| -n-N \right> \left< -N \right| \hat{H}_{n,N} \left| -n-N \right>\left< n \right| \hat{H}_n \left| 0 \right>^2}{ \left< n+N \right| \hat{H}_{n,N} \left| n+N \right>^2 \left< n \right| \hat{H}_n \left| n \right>\left< n \right| \hat{H}_n \left| 0 \right>}\right)  \label{eq10} \\ 
	\phi_{\ell \ell u}-\phi_{u \ell u}&= \arg\left(\frac{ \left< -N \right| \hat{H}_{n,N} \left| -N \right>^2\left< 0 \right| \hat{H}_n \left| 0 \right>^2}{\left< 2n+N \right| \hat{H}_{n,N} \left| 2n+N \right> \left< n+N \right| \hat{H}_{n,N} \left| 2n+N \right> \left< n \right| \hat{H}_n \left| 0 \right>\left< 0 \right| \hat{H}_n \left| 0 \right>}\right) \label{eq11} \\
	\phi_{\ell u u}-\phi_{uuu}&= \arg\left(\frac{\left< -N \right| \hat{H}_{n,N} \left| -n-N \right>^2\left< n \right| \hat{H}_n \left| 0 \right>^2}{\left< 2n+N \right| \hat{H}_{n,N} \left| n+N \right> \left< n+N \right| \hat{H}_{n,N} \left| n+N \right> \left< n \right| \hat{H}_n \left| n \right>\left< n \right| \hat{H}_n \left| 0 \right>}\right) \label{eq12}
\end{align}
\end{widetext}

\noindent
we immediately see that the matrix elements describing the first beam splitter ($\left< 0 \right| \hat{H}_n \left| 0 \right>$ in Eq.'s \ref{eq9} and \ref{eq11} and $\left< n \right| \hat{H}_n \left| 0 \right>$ in Eq.'s \ref{eq10} and \ref{eq12}) cancel out in each pair. A slightly less-ideal symmetry can be seen by looking at the left-most matrix elements in each pair, which correspond to the last beam-splitter. The contribution to the diffraction phase due to the last beam-splitter is

\begin{align*}
	\phi_{\text{last}}=\arg\Bigg(&\frac{\left< -n-N \right| \hat{H}_{n,N} \left| -N \right>}{\left< n+N \right| \hat{H}_{n,N} \left| 2n+N \right>}   \\  \times \frac{\left< n+N \right| \hat{H}_{n,N} \left| n+N \right>}{\left< -n-N \right| \hat{H}_{n,N} \left| -n-N \right>} \times
	&\frac{\left< -N \right| \hat{H}_{n,N} \left| -N \right>}{\left< 2n+N \right| \hat{H}_{n,N} \left| 2n+N \right>}  \\ \times \frac{\left< 2n+N \right| \hat{H}_{n,N} \left| n+N \right>}{\left< -N \right| \hat{H}_{n,N} \left| -n-N \right>}\Bigg)
\end{align*}

\noindent
The matrix elements that change the atom momentum (e.g. $\left< -n-N \right| \hat{H}_{n,N} \left| -N \right>$) cancel because of the symmetry \eqref{eq:diffractionsym}. Additionally, in the limit of large $N$ we satisfy Eq. \eqref{eq:undiffractedsym}, so the matrix elements that do not change momentum (e.g. $\left< n+N \right| \hat{H}_{n,N} \left| n+N \right>$) cancel out, giving $\phi_{\text{last}}=0$ under these assumptions. From these cancellations we learn two important properties:

\begin{itemize}
  \item The first beam-splitter has no effect on the measured diffraction phase in the case of a conjugate Ramsey-Bord\'e interferometer, even if the Bragg  detuning $\delta$ is not on resonance.
  \item The last beam splitter also does not contribute diffraction phase to the interferometer, assuming that $N$ is large.
\end{itemize}

Experimentally we suppress our sensitivity to the diffraction phase by varying the pulse separation time $T$ and extracting it as the $T$-independent phase (the phase containing the recoil frequency is linear in $T$, see Eq. \ref{eqn:01}). However, in practice the Bragg intensity will vary temporally within each experimental cycle, which (as will be shown in the next section) can lead to significant systematic phase shifts. The time of the 2\textsuperscript{nd} and 3\textsuperscript{rd} pulses relative to the start of the interferometer sequence can be fixed, suppressing variations in the intensity of those pulses. Since  the diffraction phase cancels exactly for the first beam splitter, the variation in the intensity of the last Bragg pulse is most important for the diffraction phase. The imperfect cancellation for the last beam splitter means that transverse motion causes suppressed effects when considering the atomic ensemble, which are discussed in the next section. 

\section{Systematic Phase  Shifts due to thermal motion of the atoms}
In the previous section we found that, to leading order, the differential diffraction phases for different output ports do not produce a systematic effect in a measurement of the fine structure constant. However, higher-order effects might produce such a systematic effect by causing the diffraction phase $\Phi$ measured by ellipse fitting to vary with the pulse separation time $T$. The next section describes a Monte Carlo simulation designed to quantify this systematic effect. 

The experiment varies the laser pulse separation time $T$ and records the laser frequency difference $\omega_m$ in the last two beam splitters needed to achieve zero interferometer phase $\Delta\Phi=0$. Using this data and rewriting (\ref{eqn:01}) in the more convenient form
\begin{align}
    \frac{\omega_m}{8(n+N)} &= \frac{\Phi}{16n(n+N)T} + \omega_r,\label{eqn:02}
\end{align}
the recoil frequency $\omega_r$ is extracted, from  which the fine-structure constant $\alpha$ may be computed. However, this is only valid if $\Phi$ remains independent of $T$, which is only approximately true. $T$-dependence of the diffraction phase can be caused by a variety of effects, and is an important systematic effect for the fine structure measurement. A linear dependence on $T$ is particularly troubling, as it would appear like a shift in $\omega_r$ and be undetectable by the experiment. 

Integrating over the ensemble introduces a dependence of the diffraction phase on the pulse separation time $T$ so that at $\Delta\Phi=0$, we can expand this systematic phase to first-order with
\begin{align}
    2n\omega_m T &= 16n(n+N)\omega_r T + \Phi_0 + \frac{d\Phi}{dT}T,
\end{align}
absorbing constant terms into $\Phi_0$ so that (\ref{eqn:02}) becomes
\begin{align}
    \frac{\omega_m}{8(n+N)} &= \frac{\Phi_0}{16n(n+N)T} + \omega_r + \frac{\frac{d\Phi}{dT}}{16n(n+N)}.
\end{align}
Thus measurements of the recoil frequency $\omega_r$ accrue a systematic error
\begin{align}
    \delta \omega_r = \frac{\frac{d\Phi}{dT}}{16n(n+N)}.
\end{align}
This results in a relative uncertainty in the fine-structure constant measurement of 
\begin{align}
    \frac{\delta\alpha}{\alpha} &= \frac{1}{2}\frac{\delta\omega_r}{\omega_r}
\end{align}
so that the relative uncertainty in $\alpha$ due to the systematic phase shift associated with thermal motion is
\begin{align}
    \frac{\delta\alpha}{\alpha} = \frac{\frac{1}{2}\frac{d\Phi}{dT}}{16n(n+N)\omega_r}.\label{eqn:ppb}
\end{align}

To determine the size of this effect, we implemented a Monte Carlo simulation in which the optical Bloch equations \eqref{eq:braggsystem} and \eqref{eq:multisystem1} for each beam in the interferometer are integrated numerically for a random sample of atoms in the thermal ensemble. The system of coupled ordinary differential equations is solved numerically for the momentum states within $[\text{min}(m,n)-20,\text{max}(m,n)+20]$ to yield the matrix elements $\braket{n|a,b|m}$, which represent amplitudes for atoms to transition from state $\ket{n}$ with momentum $2n\hbar k$ to state $\ket{m}$ after being driven by pulses with frequency differences $\omega_{12}=8a\omega_r$ and $\omega_m=8b\omega_r$, which are dependent on the velocity $v$. These matrix elements are pre-computed for various atom velocities $v$ and pulse intensity ratios $I/I_{\pi/2}$ for each set of parameters, where $I_{\pi/2}$ is the intensity needed to transfer momentum $2n\hbar k$ to atoms with 50\% probability, where $k=\frac{k_1+k_2}{2}$.

\begin{figure}[h]
    \centering
    \includegraphics[width=\columnwidth]{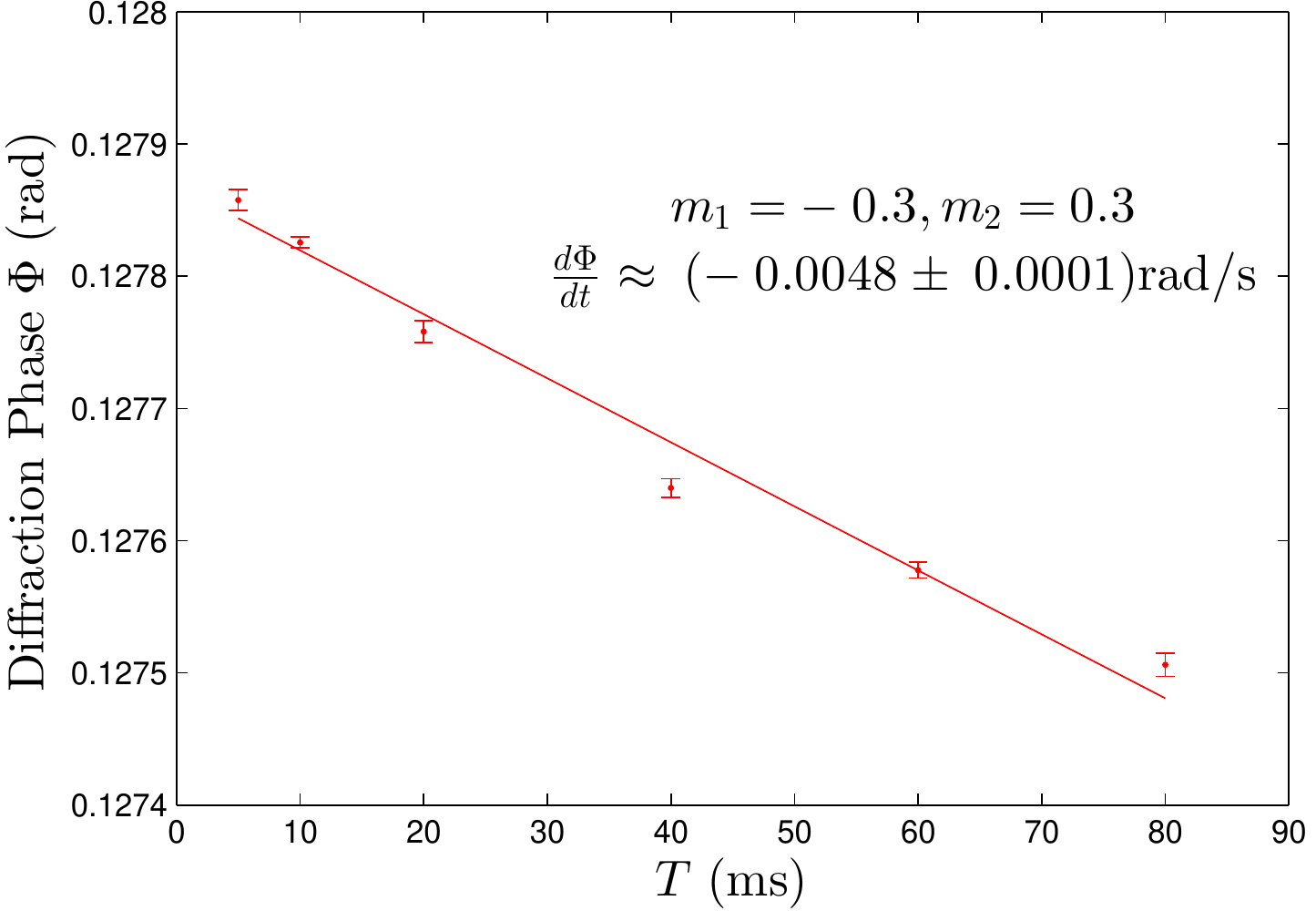}
    \caption{Simulation of the variation of the diffraction phase with $T$, taken with $10^7$ simulated atoms. $m_1$ and $m_2$ are defined in Eq. \eqref{eq:m1m2}.}
    \label{fig:fitExample2}
\end{figure}

The effect of each laser pulse beam splitter was calculated using the previously pre-computed matrix elements and interpolated over $v$ and $I/I_{\pi/2}$ using a two-dimensional complex cubic interpolation scheme. The diffraction phases for all atoms are combined with a uniform common-mode phase between 0 and $2\pi$ to generate an ellipse, which is then fit to determine the overall diffraction phase measured for that particular pulse separation time $T$. $T$ is then varied (with the time of the 2nd and 3rd pulses fixed) to determine the $T$-dependence of the diffraction phase. A typical result of such a simulation is shown in Figure \ref{fig:fitExample2}. 

A numerical determination of $\frac{d\Phi}{dT}$ can quantify the effect of the systematic phase on the uncertainty in the measurements of $\alpha$. As discussed in section \ref{theory}, the systematic shift is sensitive to the intensity of the last pulse when thermal motion is included; this dependence is shown in Figure \ref{fig:spatial_filtering}, using parameters for the Monte Carlo simulation listed in Table \ref{tab:simulate}.

\begin{table}[h]
    \centering
    \caption{Monte Carlo simulation parameters}
    \begin{tabular}{@{}ccc@{}}
        \toprule
        Name & Symbol & Value \\ \midrule
        Bragg diffraction order
        & $n$ & 5 \\
        Bloch oscillation order & $N$ & 25 \\
        Pulse width & $\tau$ & $14.5 \,\mu s$ \\
        Recoil frequency & $f_r$ & $\SI{2066}{Hz}$ \\
        Grid resolution of $(m_1,m_2)$ & $\Delta m_{1,2}$ & 0.1 \\ 
        Pulse time step &  & $0.02\tau$ \\
        Pulse time limits & $t_\text{max}$ & $\pm 3\tau$ \\
        Velocity step & $\Delta v$ & $\SI{0.02}{mm/s}$ \\
        Velocity limits & $v_\text{max}$ & $\pm\SI{0.4}{mm/s}$ \\
        Intensity ratio maximum & $r_\text{max}$ & 2 \\
        Intensity ratio step & $\Delta r$ & 0.02 \\ 
        Initial position offset & $x_{00}$ & $\SI{0}{mm}$ \\
        Spatial spread & $\sigma_{xyz}$ & $\SI{2.2}{mm}$ \\
        Horizontal velocity spread & $\sigma_{v_{xy}}$ & $\SI{1.5}{mm/s}$ \\
        Vertical velocity spread & $\sigma_{v_z}$ & $\SI{0.05}{mm/s}/\sqrt{2}$ \\
        Bragg Beam Waist &  & $\SI{6.2}{mm}$ \\
        Initial $x$-velocity & $v_{0x}$ & $\SI{0}{mm/s}$ \\
        Recoil velocity & $v_r$ & $\SI{3.522}{mm/s}$ \\
        Number of simulations & $N_\text{sim}$ & 1,000,000 \\
        4th pulse intensity ratio &  & 1 \\
        Intensity ratio & $I/I_{\pi/2}$ & 1.1 \\ 
        Pulse Separation Times & $T$ & $\left\{\text{5,10,20,40,60,80}\right\}$\,ms \\ \bottomrule
    \end{tabular}

    \label{tab:simulate}
\end{table}

\section{Spatial Filtering}
\begin{figure}[h]
    \centering
    \includegraphics[width=\columnwidth]{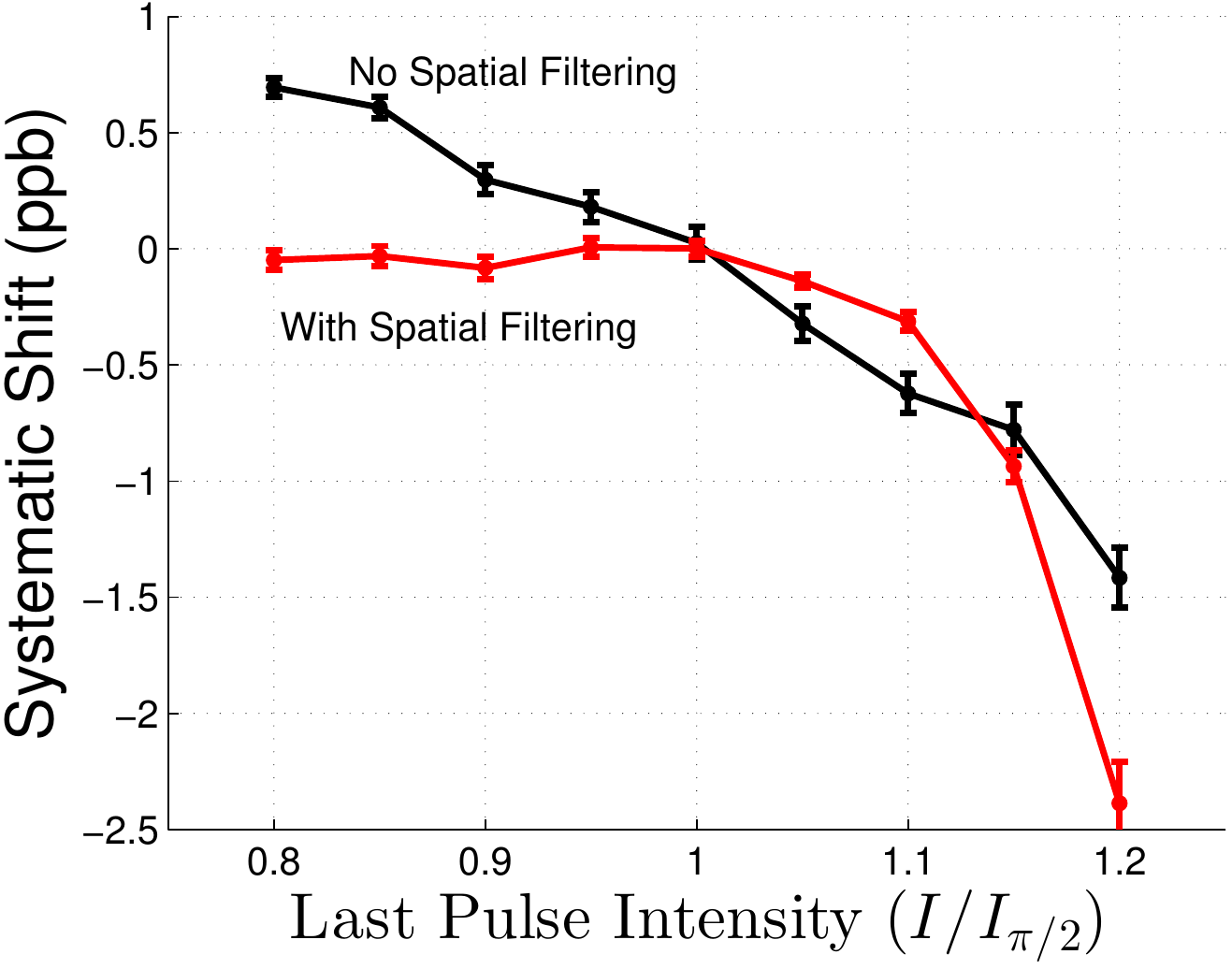}
    \caption{Simulation of the systematic shift as a function of the intensity of the last Bragg pulse, for $n$=5, $N$=25. Here, the waist of the Bragg beam is 6.2\,mm; without spatial filtering, the velocity selection beam is also 6.2\,mm. For spatial filtering, we choose 3.0\,mm in this simulation.}
    \label{fig:spatial_filtering}
\end{figure}

As the diffraction phase systematic is due to the thermal motion of the cloud, it can be suppressed with ``spatial filtering''; by controlling the detection volume and limiting the spatial extent of the atom cloud before the interferometer, only atoms in the center of the Bragg beam, with small transverse velocity, will contribute to the contrast. Experimentally, this is accomplished by using a separate beam for the velocity selection stage before the interferometer; this separate beam can have a smaller waist than the Bragg beam, keeping the Bragg intensity uniform over the atom cloud while removing atoms on the edges from the interferometer. The effect of this spatial filtering beam is shown in Figure \ref{fig:spatial_filtering}---it can be used to reduce the size of the systematic shift, at the cost of reducing the overall fluorescence signal. 

The sensitivity to the last pulse intensity is decreased when spatial filtering is used. We see that this places a constraint on the allowed variation of the last pulse intensity, but that by controlling it within 2\% we can keep the systematic shift below 0.1\,ppb. Due to the importance of the Bragg intensity for this systematic shift, the sensitivity to pulse shape is investigated in the following section. 

\section{Pulse Shape Selection}
The laser pulses used as Bragg diffraction beam splitters in the experiment are Gaussian pulses of the form
\begin{align}
    \Omega_R(t) = \hat{\Omega}_Re^{-t^2/2\tau^2}
\end{align}
with pulse width $\tau = 14.5 \,\mu s$ and truncated outside a finite time interval $t\in[-3\tau,3\tau]$ for some constant $\hat{\Omega}_R$. We have characterised the pulse intensity $I$ in terms of the corresponding Rabi frequency $\Omega_R$ which has a much smaller time-scale than the pulse width $\tau$ itself. The choice of pulse shape is not completely arbitrary; it was shown in Ref \cite{PhysRevA.77.023609} that smooth pulse shapes are optimal to avoid unwanted outputs. However, a non-Gaussian waveform can be used so long as its peak intensity can be tuned to achieve the necessary beam splitter property of transferring momentum $2n\hbar k$ to passing atoms with 50\% probability. 

A convenient basis for forming smooth non-Gaussian pulses is Hermite polynomial-corrected Gaussian pulses, which we call \emph{Hermite-Gaussian pulses}. For a constant $\hat{\Omega}_R$, a Hermite-Gaussian pulse can be constructed from the general form
\begin{align}
   \label{eq:m1m2}
    \Omega_R(t) = \hat{\Omega}_R e^{-x^2/2} [1 + m_1 H_1(x) + m_2 H_2(x) + \cdots],
\end{align}
where $x=t/\tau$ for some indicative pulse width $\tau$, while $m_1,m_2,\ldots$ are arbitrary weights and $H_j(x)$ denotes the $j$-th order Hermite polynomial in $x$ defined conventionally as
\begin{align}
    H_0(x) &= 1 \nonumber \\ 
    H_1(x) &= x\nonumber \\
    H_2(x) &= x^2 - 1 \nonumber \\
    H_3(x) &= x^3 - 3x\\
    &\vdots \nonumber
\end{align}
As the pulse profile represents an intensity, it must necessarily be positive over the time interval $t\in[-3\tau,3\tau]$. We define this set of pulse parameters as
\begin{align}
    \mathcal{D} = \{(m_1,m_2)\in\mathbb{R}^2|\Omega_R(t)\ge 0 \,\,\forall t\in[-3\tau,3\tau]\},\label{eqn:domain}
\end{align}
which forms a natural domain over which to optimize. Within this region, the Bragg pulses may be single-peaked or double-peaked. 

A linear-regression fit of the diffraction phase $\Phi$ over the pulse separation time $T$ was performed, an example of which is plotted in Figure \ref{fig:fitExample2}. The fit is used as an estimate for the systematic effect in terms of $d\Phi/dT$ from which one can estimate the effect on $\delta\alpha/\alpha$ using (\ref{eqn:ppb}). However, since the slope uncertainty can be comparable in magnitude to the slope itself, a figure of merit for $d\Phi/dT$ that also took into account $\delta\left(d\Phi/dT\right)$ was used. We define our figure of merit $\mathcal{F}_{1\sigma}(\mu,\sigma)$ to be a 68$\%$ (1-sigma) confidence upper limit, such that for a probability distribution $P(x)$ with mean $\mu$ and width $\sigma$, the metric is the number $\mathcal{F}$ so that the integral over $P(x)$ from $-\mathcal{F}$ to $+\mathcal{F}$ equals 0.68.

Using this figure of merit, we can place a reasonable bound on the systematic shift  in $\alpha$ as a relative uncertainty
\begin{align}
    \frac{\delta\alpha}{\alpha} &= \frac{\frac{1}{2}\mathcal{F}_{1\sigma}\left[\frac{d\Phi}{dT},\delta\left(\frac{d\Phi}{dT}\right)\right]}{16n(n+N)\omega_r}.\label{eqn:ppb2}
\end{align}

\begin{figure}[h]
    \centering
    \includegraphics[width=\columnwidth]{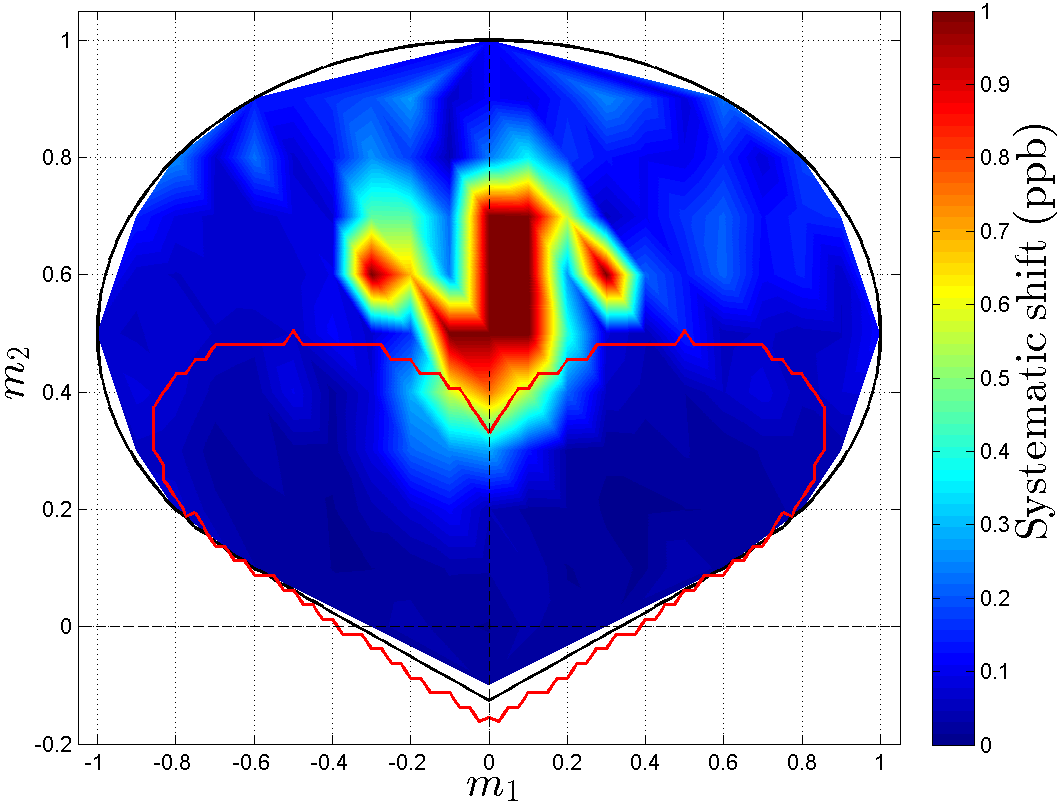}
    \caption{Colour plot of the systematic shift in units of ppb in $\alpha$ over all valid pulse shape parameters $m_1$ and $m_2$ using $N_\text{sim} = 10^6$. The colour scale is limited to $\delta\alpha/\alpha$ $\in$ [0\,ppb, 1\,ppb] and colours beyond this range have been omitted to show relevant detail. The black curve encircles the region where the intensity is always positive. The red curve encircles the region where pulses are single-peaked.}
    \label{fig:systematicppbColour1}
\end{figure}

\begin{figure}[h]
    \centering
    \includegraphics[width=\columnwidth]{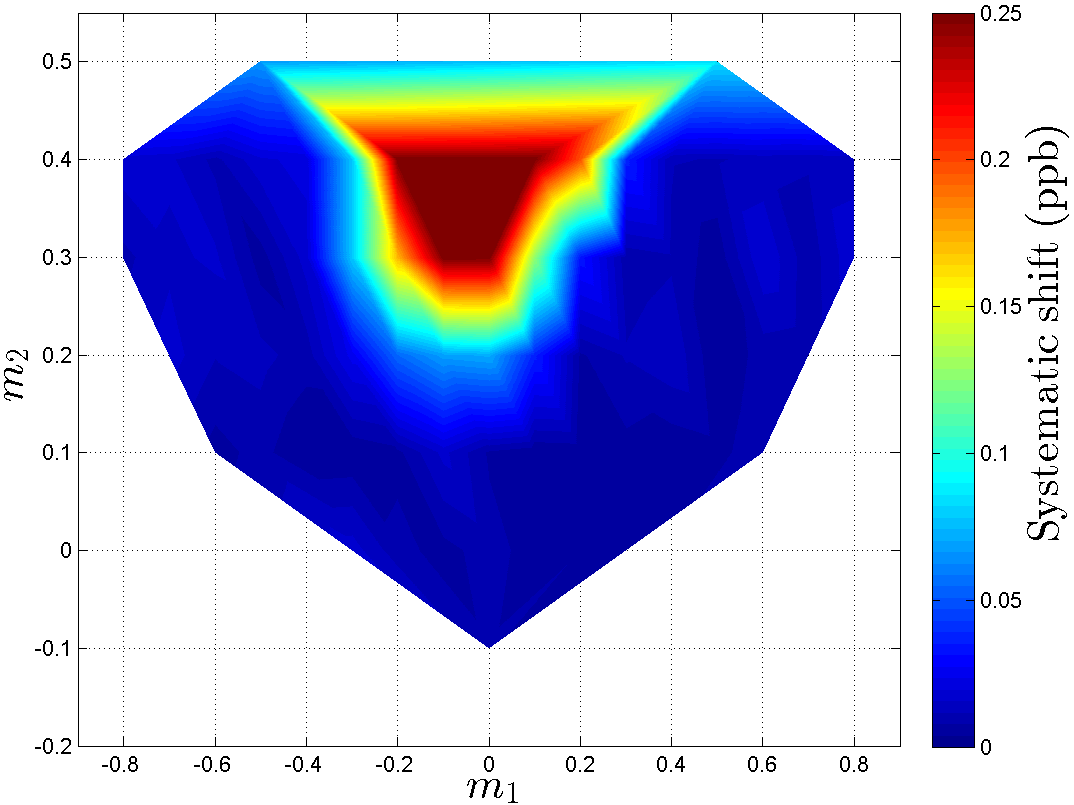}
    \caption{Color plot of the systematic shift in units of ppb in $\alpha$ with $N_\text{sim}=10^7$ over single-peak pulses. The colour scale is limited to $\delta\alpha/\alpha$ $\in$ [0\,ppb, 0.25\,ppb].}
    \label{fig:systematicppbColour2}
\end{figure}

The Monte Carlo simulation was run over the full domain $\mathcal{D}$ at $N_\text{sim}=10^6$ for each $(m_1,m_2)$ at a grid resolution of 0.1, shown in Figure \ref{fig:systematicppbColour1}. A second and finer simulation was run over the single-peak region at $N_\text{sim}=10^7$ with the same grid resolution in $(m_1,m_2)$. Figure \ref{fig:systematicppbColour2} shows the results from a longer simulation with $N_\text{sim}=10^7$ over just the single-peak region in $(m_1,m_2)$-space, and with a higher-quality slope fit as shown in Figure \ref{fig:fitExample2}. 

The figure of merit for the systematic error is well below $\SI{0.25}{ppb}$ for all single-peak pulses and is well below  $\SI{0.10}{ppb}$ for most pulses in this region. Moreover, there are even tighter bounds on the Gaussian case where the slope fit is at $\SI[separate-uncertainty=true]{0.008(3)}{ppb}$ with a figure of merit of $\SI{0.010}{ppb}$.

From the simulation results, it is apparent that Hermite-Gaussian pulses up to second order offer no advantage over Gaussian pulses in suppressing the systematic error, especially those pulses which had two peaks. Furthermore, within the regime of single-peaked pulses, the systematic error was typically less than $\SI{0.1}{ppb}$ for near-Gaussian peaks as summarised in Figure \ref{fig:systematicppbColour2}. We can put an upper bound on the systematic shift in the Gaussian case of about $\SI{0.008}{ppb}$ using our figure of merit at $N_\text{sim}=10,000,000$. An even finer simulation for the Gaussian case may be able to resolve a decisively non-zero systematic shift. From the studies done thus far, Gaussian pulses are indeed the best. 

\begin{figure}[h]
    \centering
    \includegraphics[width=\columnwidth]{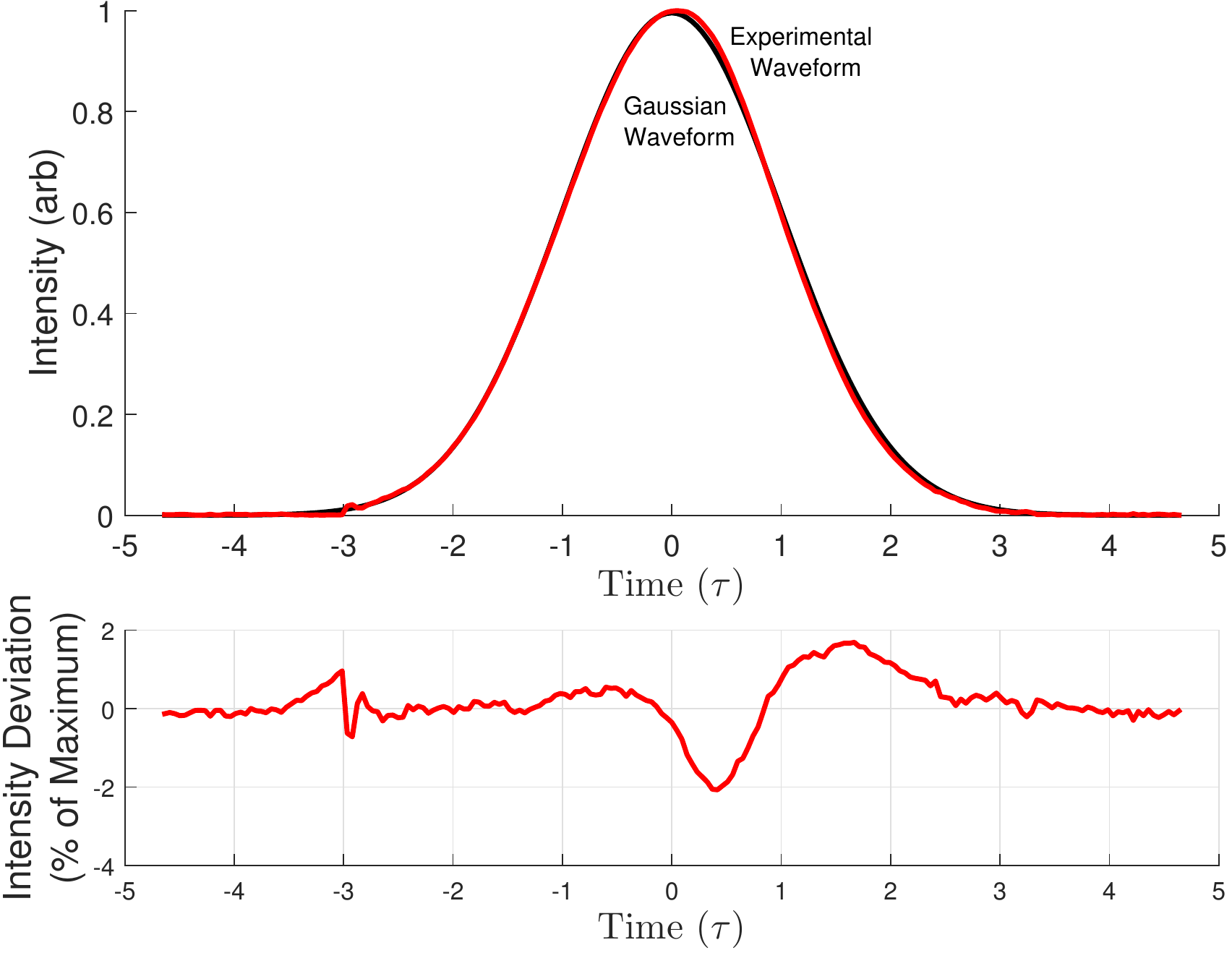}
    \caption{Comparison of the Bragg waveform after the intensity servo (red) vs. an ideal gaussian (black). The two waveforms are on top; the difference between the two is on the bottom. Note the truncation of the waveform around -3$\tau$ and 3$\tau$. The deviation can be reduced by suitable adjustment of the servo; however, even this deviation results in a negligible shift of less than 0.03 ppb.}
    \label{fig:waveform}
\end{figure}

The waveform we use in the actual experiment is determined by an intensity servo which locks our pulse shape to a reference waveform; the resulting waveform will not match a perfect Gaussian (see Figure \ref{fig:waveform}). We can therefore measure the experimental waveform and run the Monte Carlo with that pulse shape. This analysis results in a systematic shift in $\alpha$ below 0.03\,ppb. 

\section{Parasitic Interferometers}

Because Bragg diffraction populates more than the two desired momentum states, it is possible to create unwanted Ramsey-Bordé interferometers \cite{1367-2630-15-2-023009}. These interferometers will close at the same time as the main interferometer, and will not be suppressed by the Bloch pulse, because they travel at the same velocity as the main interferometer. The effect of a parasitic path of order $n_{p}$ on the outputs $\tilde{\Psi}_{ij}$ of a Ramsey-Bordé interferometer of order $n$ is:

\begin{figure}[h]
	\centering
	\includegraphics[width=\columnwidth]{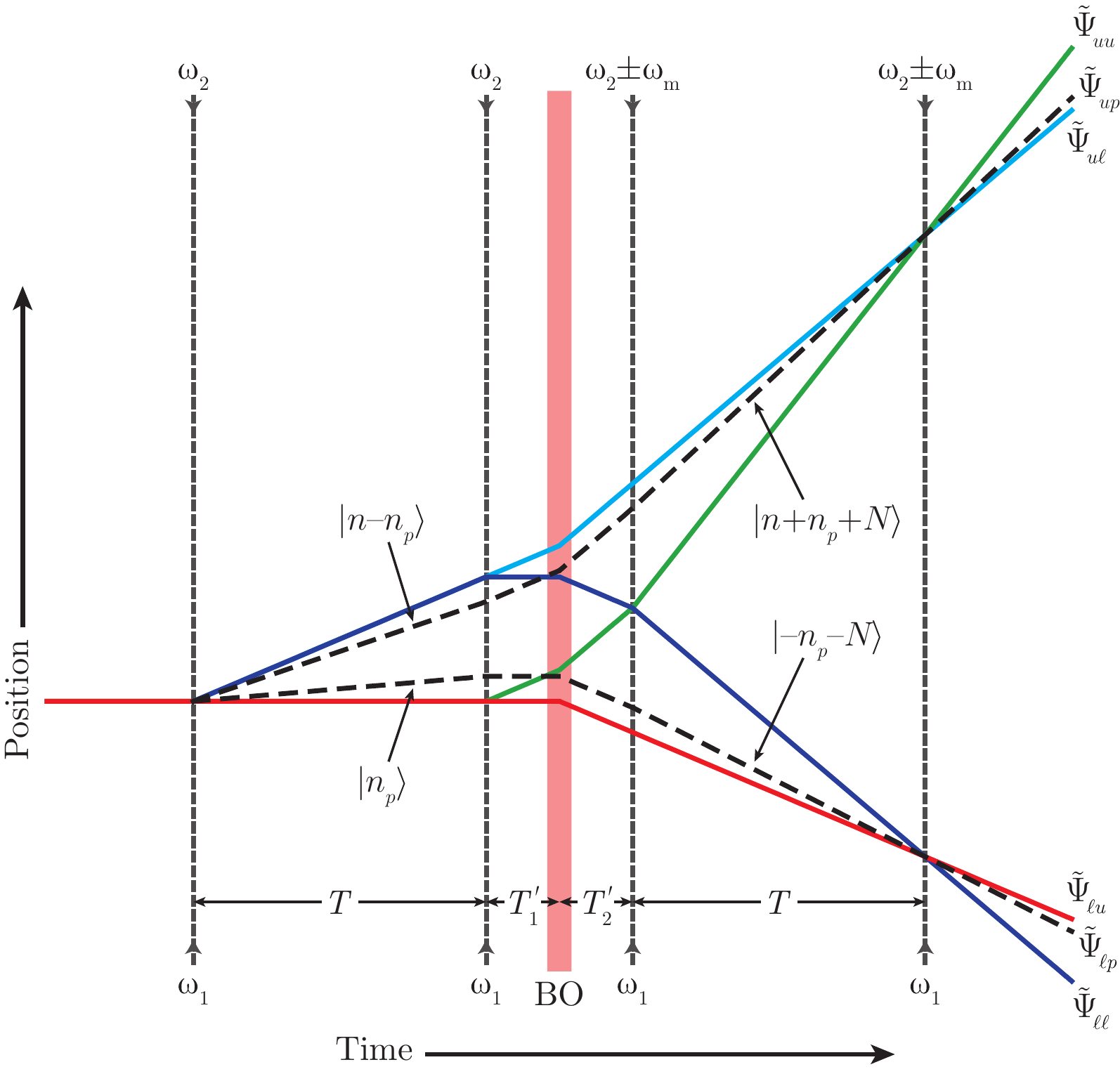}
	\caption{Geometry of the interferometers when parasitic Ramsey-Bordé interferometers (dashed lines) are included.}
	\label{fig:rbmomentumparasitic}
\end{figure}

\begin{align*}
\lvert \tilde{\Psi}_{uu} \rvert^{2} &= \lvert  c_{uuu} + c_{u \ell u} e^{-i \phi_{c}} + c_{upu} e^{i n_{p} (n_{p}-n) \omega_{r} T -i n_{p}\phi_{c}/n  } \rvert^{2}, \\
\lvert \tilde{\Psi}_{u \ell} \rvert^{2} &= \lvert  c_{uu \ell} + c_{u \ell \ell} e^{-i \phi_{c}} + c_{up \ell} e^{i n_{p} (n_{p}-n) \omega_{r} T -i n_{p}\phi_{c}/n } \rvert^{2}, \\
\lvert \tilde{\Psi}_{\ell u} \rvert^{2} &= \lvert  c_{\ell uu} e^{i \phi_{c}} + c_{\ell \ell u}  + c_{\ell pu} e^{i n_{p} (n_{p}-n) \omega_{r} T +i n_{p}\phi_{c}/n } \rvert^{2}, \\
\lvert \tilde{\Psi}_{\ell \ell} \rvert^{2}&= \lvert  c_{\ell u \ell} e^{i \phi_{c}} + c_{\ell \ell \ell}  + c_{\ell p \ell} e^{i n_{p} (n_{p}-n) \omega_{r} T +i n_{p}\phi_{c}/n } \rvert^{2}, 
\end{align*}
where $\phi_{c}$ is the common-mode phase due to accelerations, and the parasitic complex amplitudes are

\begin{align*}
c_{upu} &=  \left< 2n+N \right| \hat{H}_{n,N} \left| n+n_{p}+N \right> &\\ &\times \left< n+n_{p}+N \right| \hat{H}_{n,N} \left| n+N \right> &\\ &\times \left< n \right| \hat{H}_{n} \left| n-n_{p} \right>\left<  n-n_{p} \right| \hat{H}_{n} \left| 0 \right>, &\\
c_{up \ell} &= \left< n+N \right| \hat{H}_{n,N} \left| n+n_{p}+N \right> &\\ &\times \left< n+n_{p}+N \right| \hat{H}_{n,N} \left| n+N \right> &\\ &\times \left< n \right| \hat{H}_{n} \left| n-n_{p} \right>\left< n-n_{p} \right| \hat{H}_{n} \left| 0 \right>, &\\
c_{\ell pu} &= \left< -N \right| \hat{H}_{n,N} \left| -n_{p}-N \right>^2  \left< 0 \right| \hat{H}_{n} \left| n_{p} \right>^{2}, &\\
c_{\ell p \ell} &= \left< -n-N \right| \hat{H}_{n,N} \left| -n_{p}-N \right> \left< -n_{p}-N \right| \hat{H}_{n,N} \left| -N \right> &\\ & \times \left< 0 \right| \hat{H}_{n} \left| n_{p} \right>^{2},
\end{align*}
with the parasitic paths shown in Figure \ref{fig:rbmomentumparasitic}. As before, the coefficients $c_{ijk}$ are calculated numerically by the Monte Carlo simulation. The two ports $\tilde{\Psi}_{up}$ and $\tilde{\Psi}_{lp}$ can also be included, but are not significantly populated and therefore do not change the results. The common-mode phase $\phi_{c}$ is modeled as a gaussian distribution, with a spread produced by vibrations of the retroreflecting mirror \cite{:/content/aip/journal/rsi/70/6/10.1063/1.1149838}. 

\begin{figure}[h]
    \centering
    \includegraphics[width=\columnwidth]{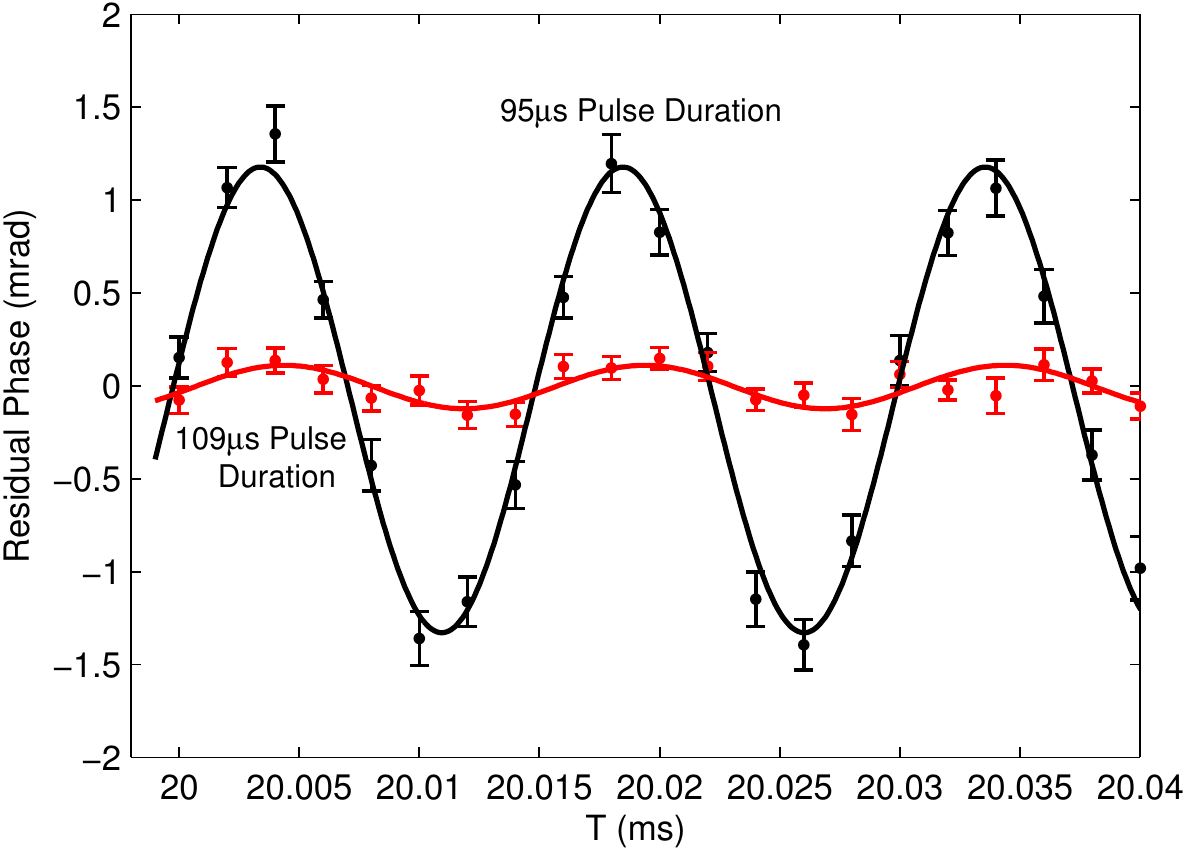}
    \caption{Monte Carlo simulation with Bragg durations of 95\,$\mu s$ and 109\,$\mu s$. The effect of the parasitic interferometer is to produce a high-frequency oscillating phase, clearly visible when using a 95$\mu s $ pulse. The parasitic interferometer is suppressed with the proper choice of pulse duration. For this simulation, the overall Bragg intensity $I/I_{\pi/2} = 1.0$, and the common-mode phase is a gaussian with width 6\,rad and offset 4\,rad. The solid lines are sinusoidal fits (all parameters free).}
    \label{fig:ParasiticSimulation}
\end{figure}

The effect of a parasitic interferometer is to produce a phase shift that oscillates in $T$ at high frequency, as in Figure \ref{fig:ParasiticSimulation}. This can produce a systematic shift in a measurement of $\alpha$ as the Ts chosen will alias this oscillation, in general not averaging out the phase shifts. For example, with typical experimental parameters an 8\,mrad peak-to-peak oscillation can produce a systematic shift in the measured $\alpha$ as large as 1\,ppb. 

\begin{figure}[t]
    \centering
    \includegraphics[width=\columnwidth]{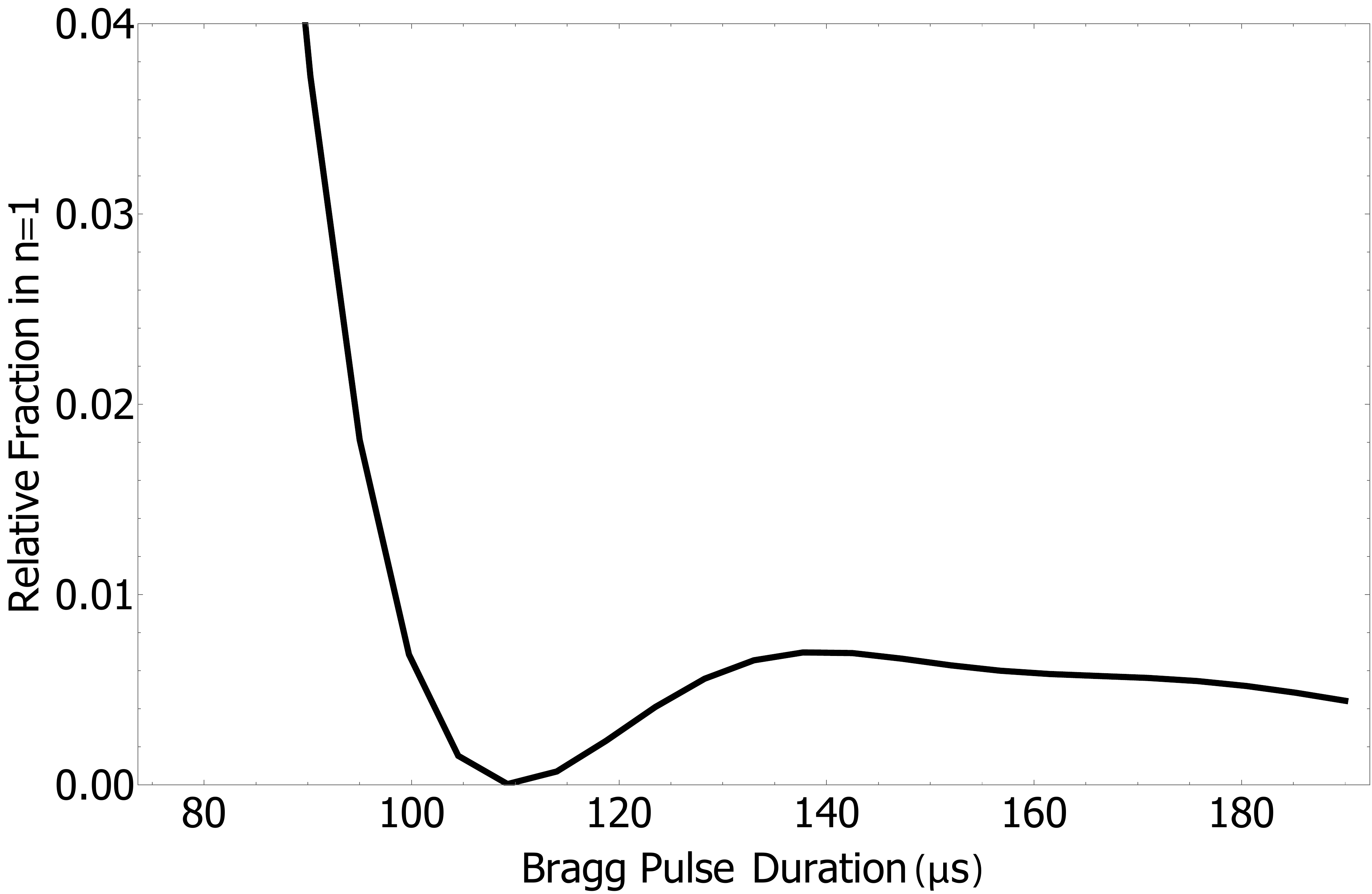}
    \caption{Fraction of atoms driven into $n$=1 as the Bragg duration is varied. This is a single-atom simulation, with the atom on Bragg resonance.}
    \label{fig:diffraction_phase_simplified}
\end{figure}

For an $n$=5 Ramsey-Bordé interferometer using $\sigma = 14.5 \,\mu s$, the dominant parasitic interferometer is $n_{p}$=1. The predicted oscillation frequency is therefore $8 n_{p} (n-n_{p}) \omega_{r}/2 \pi=66\,$kHz. By taking multiple pairs of Ts, with each pair separated by half the period of the oscillation, we can dramatically suppress the sensitivity to this effect by averaging over it, limited by the uncertainty in the period. Even with 5\% error in the period, using pairs of $T'$s reduces the systematic shift for an 8\,mrad oscillation to less than 0.1\,ppb. 

\begin{figure}[h]
    \centering
    \includegraphics[width=\columnwidth]{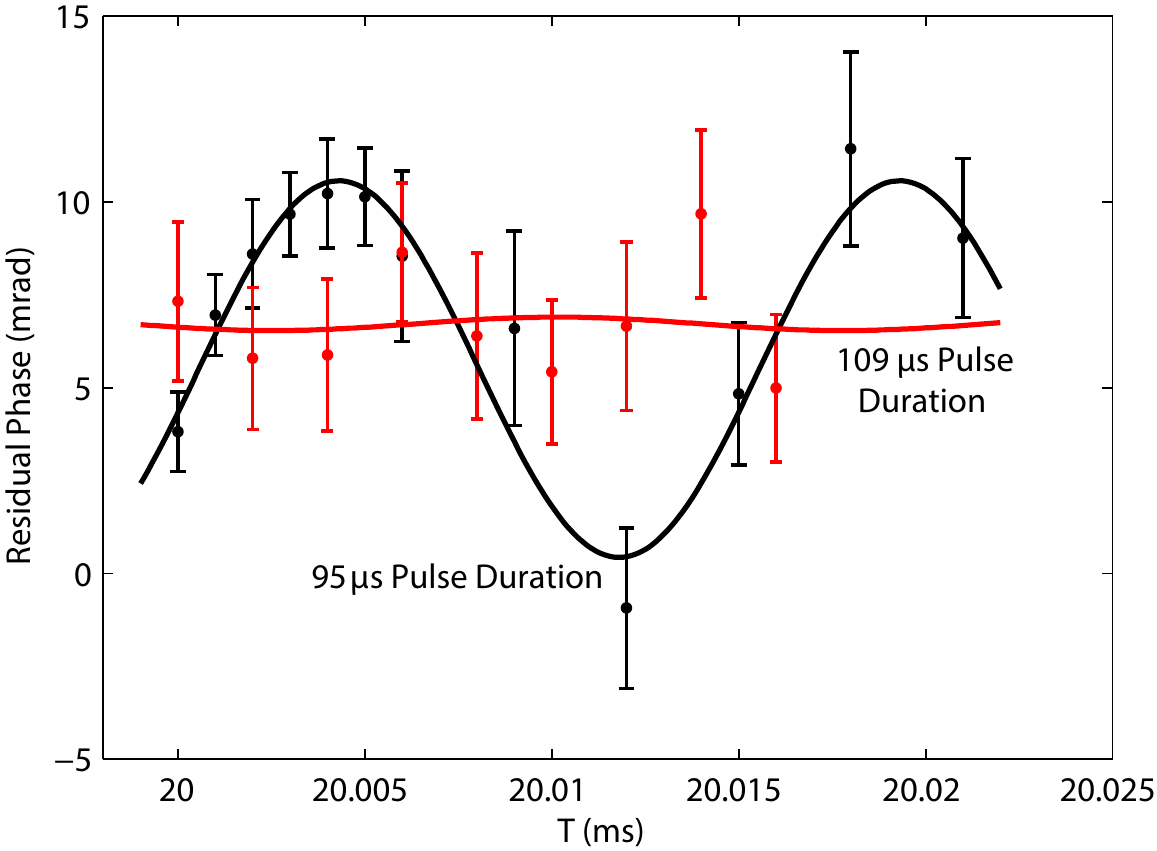}
    \caption{Experimental comparison between Bragg durations of 95\,$\mu s$ and 109\,$\mu s$. The solid lines are sinusoidal fits (all parameters free). The presence of a parasitic interferometer is clearly visible when 95\,$\mu s$ is used, and it is suppressed when 109\,$\mu s$ is used instead.}
    \label{fig:ParasiticData}
\end{figure}

The population driven into an undesired order depends sensitively on the Bragg intensity and detuning from Bragg resonance; therefore, the amplitude for an ensemble can be much larger than that predicted by a na\"{i}ve calculation in which the atom is on Bragg resonance. However, such a calculation predicts a ``magic'' duration, see Figure \ref{fig:diffraction_phase_simplified}, which minimizes this population and suppresses the parasitic interferometer, without sacrificing the interferometer contrast. A more complete Monte Carlo simulation is shown in Figure \ref{fig:ParasiticSimulation}, which demonstrates suppression in the phase shift due to a parasitic interferometer by an appropriate choice of Bragg pulse duration. 

Experimental data with the same parameters is shown in Figure \ref{fig:ParasiticData}. A sinusoidal fit to the 95\,$\mu s$ data with no fixed parameters shows an $8 \pm 1$\, mrad peak-to-peak oscillation. The fitted frequency is $65.99 \pm 0.03$\,kHz, in agreement with the expected 66\,kHz expected for $n_{p}$=1. A Bragg pulse duration of 109\,$\mu s$ yields no observable oscillation. The amplitude of any residual oscillation is constrained to be less than 2.6\,mrad (p-value <0.05), limited by statistics. It is important to note that this model can predict the frequency and phase of the parasitic oscillation observed in the experiment, but cannot accurately predict the amplitude of the oscillation, which will depend sensitively on the overall Bragg intensity, the intensity of the last Bragg pulse, the 2-photon detuning, and the vibration spectrum experienced by the retroreflection mirror. Fortunately, the model can be used as a guide to suppress the effect. 

\begin{table}[t]
    \centering
    \caption{Systematic shifts from ellipse distortion due to the multi-port nature of Bragg diffraction. First we consider the systematic shift due to the atoms' thermal motion when using spatial filtering, then allow parameters to vary, replicating the level of control achieved in the actual experiment. We also consider the deviation from perfect Gaussian we achieve for the Bragg pulse waveform when using our intensity servo, as well as the influence of parasitic interferometers when operating at the ``magic'' pulse duration and when using half-period $T'$s.}
    \begin{tabular}{@{}ccc@{}}
        \toprule
        Effect & Value & $\delta \alpha/\alpha$ (ppb) \\ \midrule
        Thermal motion of atoms\\ without parameter variations & N/A & $\pm 0.02$ \\ \midrule
        Cloud radius [mm] & $2.2 \pm 1$ & $\pm 0.03$ \\
        Vertical velocity width [$v_r$] & $1.5 \pm 0.25$ & $\pm 0.06$ \\
        Ensemble hor. velocity [$v_r$] & $0 \pm 0.5$ & $\pm 0.01$ \\
        Initial hor. position [mm] & $0 \pm 1$ & $\pm 0.04$ \\ 
        Intensity [$I_{\pi/2}$] & $1.05 \pm 0.02$ & $\pm 0.04$ \\ 
        Last pulse intensity ratio &  $1.0 \pm 0.02$  & $\pm 0.03$ \\ \midrule
        Non-gaussian waveform \\ (used in actual experiment) & Figure \ref{fig:waveform} & $\pm 0.03$ \\ \midrule
        Parasitic Interferometers & <2.6\,mrad  & $\pm 0.03$ \\ \midrule
        Total &  & $\pm 0.10$ \\ \bottomrule
    \end{tabular}
    \label{tab:Conclusion}
\end{table}

\section{Conclusion}
The results of the systematics analysis in this paper are summarized in Table \ref{tab:Conclusion}. The differential diffraction phase between output ports of a conjugate Ramsey-Bord{\'e} interferometer does not produce a systematic effect to leading order, but has the potential to create large systematic errors when the thermal motion of the atoms is included, limiting the precision of the measurement of the fine structure constant. A simulation incorporating Bragg diffraction-based beam splitters has identified a range of parameters for which the systematic effects discussed in this paper can be suppressed to below 0.1\,ppb. Second-order Hermite-Gaussian pulses were studied and found to offer no benefit over pure Gaussian pulses, confirming that Gaussian pulses are the best (among those studied). The use of Bragg diffraction results in parasitic interferometers which can be controlled, by a proper choice of the Bragg pulse separation and by use of a ``magic'' pulse duration. Enhancements to the contrast of the interferometer, by means of Stark compensation, allow direct comparison between simulation and experiment---and yield the largest matter wave phase difference ever created or measured in an atom interferometer of any kind with a nonzero free-evolution phase.

\section*{Acknowledgements}
We would like to thank Jiafeng Cui and Xuejian Wu for their assistance in developing the Stark compensation beam. This material is based upon work supported by the National Science Foundation under CAREER Grant No. PHY-1056620, the David and Lucile Packard Foundation, and National Aeronautics and Space Administration Grants No. NNH13ZTT002N, No. NNH10ZDA001N-PIDDP, and No. NNH11ZTT001. We also acknowledge collaboration with Honeywell Aerospace under DARPA contract No. W31P4Q-13-C-0092. 


\bibliographystyle{unsrt}

\end{document}